\documentclass[prl,twocolumn,superscriptaddress]{revtex4}
\usepackage{graphicx}
\usepackage{amsfonts}
\usepackage{amsmath}
\usepackage{amssymb}  
\usepackage{bm}
\usepackage{hyperref}
\usepackage{slashed}
\usepackage{color}
\begin{document}
\title{Theory of competing Chern-Simons orders and emergent phase transitions}
\author{Rui Wang}
\affiliation{National Laboratory of Solid State Microstructures and Department of Physics, Nanjing University, Nanjing 210093, China}
\affiliation{Collaborative Innovation Center for Advanced Microstructures, Nanjing 210093, China}
\author{Z. Y. Xie}
\email{qingtaoxie@ruc.edu.cn}
\affiliation{Department of Physics, Renmin University of China, Beijing 100872, China}
\author{Baigeng Wang}
\email{bgwang@nju.edu.cn}
\affiliation{National Laboratory of Solid State Microstructures and Department of Physics, Nanjing University, Nanjing 210093, China}
\affiliation{Collaborative Innovation Center for Advanced Microstructures, Nanjing 210093, China}
\author{Tigran Sedrakyan}
\affiliation{Department of Physics, University of Massachusetts Amherst, Amherst, Massachusetts 01003, USA}
\begin{abstract}
Based on the Chern-Simons fermionization of spin-$1/2$ operators, we propose a systematic framework to investigate the competition between emergent phases in frustrated two-dimensional XY
quantum magnets. Application of the method to the antiferromagnetic honeycomb 
spin-$1/2$ $J_1$-$J_2$ XY model reveals an unconventional phase transition between two Chern-Simons orders: the Chern-Simons superconductor and the exciton insulator of Chern-Simons fermions. We show that in the spin language, this transition translates to the transition from the planar N\'{e}el state to the non-uniform chiral spin-liquid that was proposed earlier in the literature.  Namely,  the Chern-Simons superconductor describes the planar N\'{e}el state, while the Chern-Simons exciton insulator corresponds to the non-uniform chiral spin-liquid. These results are further confirmed by our high-precision tensor network calculation, which provides the first numerical evidence for the transition from N\'{e}el order to a non-uniform chiral spin-liquid. We argue that the developed method can be applied to other frustrated quantum magnets of XXZ type and can detect unconventional phase transitions.


\end{abstract}

\maketitle

\emph{\color{blue}{Introduction.--}}
Landau's concept of spontaneous symmetry breaking provides a universal paradigm for conventional phase transitions. However, recent years have witnessed the discovery of many novel states beyond the Landau' paradigm, including the fractional quantum Hall effects \cite{Klitzing,R. B. Laughlin,D C Tsui,F D M Haldane} and quantum spin-liquids \cite{PWAnderson,LBalents,palee,xgwena}. Some of these findings evoked great interest in topological orders (TOs) \cite{xgwenc}, featured by their ground state degeneracy \cite{xgwenb} and the emergence of anyons \cite{fwilczek,akitaev}.
An important but unsolved issue is to theoretically understand the unconventional phase transitions (UPTs) \cite{tsenthil,mlevinb,avishwanath,tsenthilb,anders,adam} from the symmetry-broken phases to the TOs \cite{yangqi,cenkexu,alex}, which are complicated by the drastic change of the systems' quantum entanglements \cite{xchen,mlevin,akitaevb,huili}. Among them, the UPTs from the N\'{e}el antiferromagnetic (AFM) order to the chiral spin-liquid (CSL) \cite{vkalmeyerr} have attracted substantial attention, because they were conjectured to have interplay with a variety of exotic phenomena, including the emergent high-$T_c$ superconductivity \cite{Konika,rbLaughlin,kyyang,Rokhsar,Kotliar,Chatterjee}.

The CSLs are fractionalized states that break time-reversal ($\mathcal{T}$), parity ($\mathcal{P}$) symmetry, resembling the fractional quantum Hall states \cite{vkalmeyerr}. In stark contrast, the N\'{e}el AFM enjoys the long-range magnetic ordering with the spin-wave excitations and has no fractionalization. Because of their disparate physical natures, the unified description of the two types of phases, which is highly desired, is still missing.  This inevitably hinders the development of a unified theory for the UPTs between them.

\begin{figure}[htb]\label{fig1} 
\includegraphics[width=\linewidth]{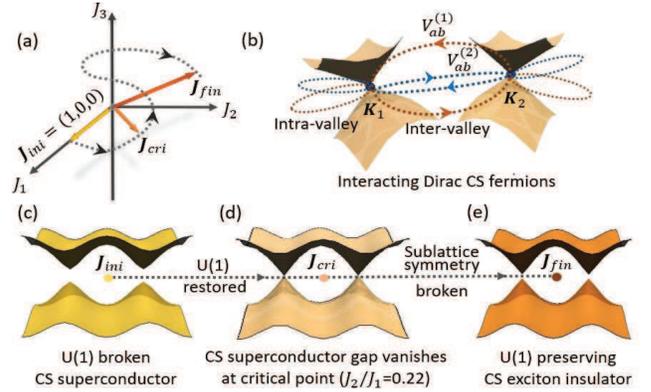}
\caption{(a) Parameters tuned from the frustration-free  $\mathbf{J}_{ini}$ to the strong-frustration $\mathbf{J}_{fin}$. (b) $H_{XY}$ is mapped in low-energy to Dirac CS fermions interacting via multiple inter- and intra-valley interactions.  (c),(d),(e) illustrate the competition of CS orders for the $J_1$-$J_2$ XY  honeycomb model, with increasing frustration.  }
\end{figure}

In this Letter, we present a framework to analyze the possibility of having UPTs in frustrated two-dimensional (2D) XXZ quantum magnets, where the symmetry-broken phases and the fractionalized states are unified within the same physical picture, namely the theory of interacting Dirac Chern-Simons (CS) fermions \cite{Tigrana,Ruia,ruib,triangular,kagome,MS2,SGK2,ruid,SGK1}, as indicated by Fig.1(b). Within this framework, we then develop an analytical theory of the UPT from the N\'{e}el AFM to a non-uniform CSL (a CSL coexists with an out-of-plane N\'{e}el order \cite{SGK2}). This is demonstrated by using a concrete example, namely the spin-$1/2$ $J_1$-$J_2$ XY honeycomb model. 
We firstly present the state-of-the-art numerical tensor-network-based calculation \cite{zyxiea,bxzheng,Corboza,Corbozb,lwanga,hjliaoa}, which suggests an interesting UPT after the N\'{e}el AFM is destablized. 
We then show that this UPT can be understood as a transition between two competing CS orders in this model, one being the $\mathrm{U}(1)$-broken CS superconductor \cite{Ruia} as shown by Fig.1(c), and the other a $\mathrm{U}(1)$-preserving CS exciton insulator (EI) indicated by Fig.1(e). With increasing frustration, the competition between the two CS orders becomes most fierce and drives the system into the criticality at $J_2/J_1\simeq0.22$ with spontaneous $\mathcal{T}$, $\mathcal{P}$-breaking. These findings better clarify and are in mutual support with our high-precision tensor network results. The latter identifies supportive evidences for a transition (at $J_2/J_1\simeq0.22$) into a $\mathcal{T}$, $\mathcal{P}$-broken phase, which displays simultaneously an out-of-plane N\'{e}el order and an entanglement spectrum consistent with $\mathrm{SU}(2)_1$ Wess-Zumino-Witten (WZW) theory \cite{WZW}. 


\emph{\color{blue}{Model and simulation.--}}
We start with a general Hamiltonian of the 2D quantum XY model on a generic lattice
\begin{equation}\label{eq1}
  H_{XY}=\sum_{\mathbf{r},\mathbf{r}^{\prime}}(J_{\mathbf{r},\mathbf{r}^{\prime}}/2)(\hat{S}^+_{\mathbf{r}}\hat{S}^-_{\mathbf{r}^{\prime}}+\hat{S}^-_{\mathbf{r}}\hat{S}^+_{\mathbf{r}^{\prime}}),
\end{equation}
where $\hat{S}^{\pm}_{\mathbf{r}}$ is the spin-raising/lowering operator at site $\mathbf{r}$. Interaction parameters $J_{\mathbf{r},\mathbf{r}^{\prime}}>0$ involve couplings up to the $n$-th nearest neighbour, forming an $n$-dimensional vector $\mathbf{J}=(J_1,J_2,...,J_n)$. With tuning  $\mathbf{J}$, we suppose that the system is initially frustration-free at $\mathbf{J}_{ini}$, and evolves into strong-frustration at $\mathbf{J}_{fin}$, as schematically shown in Fig.1(a).
A simple choice for $\mathbf{J}_{ini}$ is $\mathbf{J}_{ini}=(1,0,...,0)$ on a bipartite lattice, for which the ground state is known to be the planar N\'{e}el AFM \cite{kennedy,dhlee}. 

We firstly present our numerical results for the aforementioned $J_1$-$J_2$ XY model \cite{zzhu,cnvarney,Carrasquilla,plekhanov,yixuanhuang} with $\mathbf{J}=(J_1,J_2)$
on honeycomb lattice, obtained by employing the tensor-network algorithm \cite{sup}.  With increasing $J_2/J_1$, an  intermediate state with an unexpected out-of-plane Ising N\'{e}el order was found by previous studies \cite{cnvarney,Carrasquilla,zzhu,plekhanov,yixuanhuang,SGK2}, which was proposed as a manifestation of a non-uniform CSL \cite{SGK2}.  However, the key numerical evidences for $\mathcal{T}$, $\mathcal{P}$-breaking in thermodynamic limit as well as the chiral edge state have been missing. Thus, the nature of the intermediate phase, and more importantly, the mechanism of the phase transition remains unclear.


To be specific, we use imaginary-time evolution to determine the ground state that is represented as the projected entangled simplex state (PESS) ansatz \cite{PESS2014, PEPS2004}, and then we evaluate the magnetic order and chirality order (defined as $\chi=\langle\hat{S}_{\mathbf{r}_1}\times(\hat{S}_{\mathbf{r}_2}\cdot\hat{S}_{\mathbf{r}_3})\rangle$) from the representation. In order to visualize the chirality more clearly, a $\mathcal{T}$-breaking perturbation seed $\beta\sim 10^{-4}$ is introduced in the calculation. Fig.2 shows the results for the bond dimension D = 14. As shown by Fig.2(a), an intermediate phase in about $J_2/J_1\sim(0.22,0.33)$ is found, which is disordered in xy-plane but has out-of-plane N\'{e}el ordering \cite{zzhu}.
Interestingly, as shown by Fig.2(b), we observe a significant enhancement of the chirality for $J_2/J_1\sim(0.22,0.33)$, indicating $\mathcal{T}$, $\mathcal{P}$-breaking of the intermediate phase. The entanglement spectrum is shown by Fig.2(c), where a nontrivial degeneracy $1,1,2,3,5,...$ is  obtained \cite{huili}, in consistent with  $\mathrm{SU}(2)_1$ Wess-Zumino-Witten conformal field theory \cite{WZW}, suggesting the existence of chiral edge state.
These results indicate that an UPT takes place at $J_2/J_1\simeq0.22$ from the planar N\'{e}el AFM to some topological phase, which  simultaneously exhibits the out-of-plane N\'{e}el order and the chiral edge state.
 
\begin{figure}[t]\label{fig2}
\includegraphics[width=\linewidth]{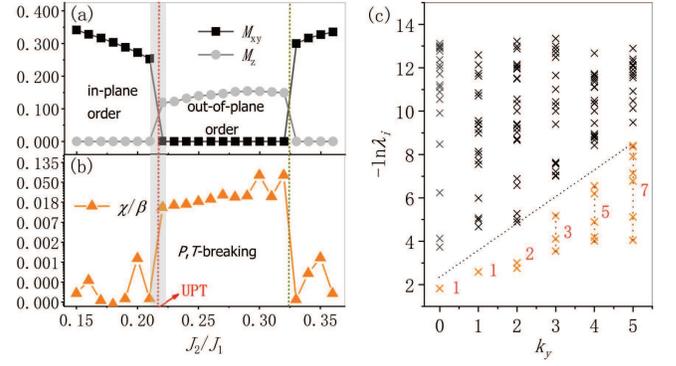}
\caption{The tensor network results on the magnetization and entanglement spectrum for $D=14$. (a) The planar N\'{e}el AFM order is stable for $J_2/J_1\lesssim0.22$ with nonzero $M_{xy}$ defined by $M_{xy}=\sum_{\mathbf{r}}\sqrt{M^2_{\mathbf{r},x}+M^2_{\mathbf{r},y}}/N$. After crossing the UPT (the shaded region), an intermediate state is found with out-of-plane Ising N\'{e}el order, i.e., $M_{z}=\sum_{\mathbf{r}}|{M_{\mathbf{r},z}}|/N\neq0$. A subsequent phase transition takes place at larger $J_2/J_1$,  which is not the focus of this work. (b) The chirality susceptibility undergoes a significant jump after the UPT. (c) The entanglement spectrum obtained from a cut in an infinite cylinder with 6 unit cells along $L_y$. $J_2/J_1=0.3$ and $k_y$ is in unit of $\pi/3$. }
\end{figure}

\emph{\color{blue}{Theoretical framework.--}}
We now start with the general XY model in Eq.\eqref{eq1}. The spin-1/2 operators, $\hat{S}^{\pm}_{\mathbf{r}}$, are algebraically equivalent to hardcore bosons and can be conveniently represented by \cite{SGK1,SGK2,Ruia,Tigrana} $\hat{S}^{\pm}_{\mathbf{r}}= f^{\pm}_{\mathbf{r}} U^{\pm}_{\mathbf{r}}$, where $ f^{\pm}_{\mathbf{r}}$ are creation/annihilation operators of the spinless CS fermions,  $ U^{\pm}_{\mathbf{r}}=\mathrm{exp}[\pm ie\sum_{\mathbf{r}^{\prime}\neq\mathbf{r}}\mathrm{arg}(\mathbf{r}-\mathbf{r}^{\prime}) n_{\mathbf{r}^{\prime}}]$ are the fermion string operators with $ n_{\mathbf{r}}$ representing the particle number operator at site $\mathbf{r}$, and $e$ being the CS charge that takes value of odd integers \cite{SGK2,Ruia,Tigrana}.

For the frustration-free point $\mathbf{J}_{ini}$, we apply the CS fermionization and work with the fermion representation of the model. The corresponding effective action can be generally derived in the long-wavelength regime as \cite{sup,Ruia},\begin{equation}\label{eq3}
\begin{split}
  S_{XY}&=\sum^{N}_{a=1}\int d^3rf^{\dagger}_{a,r} i\slashed{D}f_{a,r}+\frac{1}{4\pi}\int d^3r\epsilon^{\mu\nu\rho}A_{\mu}\partial_{\nu}A_{\rho},
\end{split}
\end{equation}
where the CS fermions enjoying linear dispersion and pseudospin (sublattice) are coupled to the CS gauge field $A_{\mu}$ generated by the string operators $U^{\pm}_{\mathbf{r}}$.  $r=(t,\mathbf{r})$, $\slashed{D}=\sigma^{\mu}(i\partial_{\mu}-eA_{\mu})$, with $\mu=0,1,2$, and $\sigma_{\mu}$ is the Pauli matrix defined in the $A,B$ sublattice space.  $f_{a}=[f_{a,A,r},f_{a,B,r}]^T$ is a Dirac spinor, where $a=1,2,...,N$ denotes the N Dirac valleys located at the momentum $\mathbf{K}_a$ \cite{Ruia}.  Further integrating out the CS gauge field $A_{\mu}$, a nonlocal interaction between the Dirac CS fermions is generated as,
\begin{equation}\label{eq4}
  V^{(1)}_{int}=\int d^3r V^{(1)\alpha\beta\rho\sigma}_{ab}(\mathbf{r}-\mathbf{r}^{\prime})f^{\dagger}_{a,\alpha,r}f_{a,\beta,r}f^{\dagger}_{b,\rho,r^{\prime}}f_{b,\sigma,r^{\prime}},
\end{equation}
where and hereafter the repeated indices are summed. The vertex  $V^{(1)\alpha\beta\rho\sigma}_{ab}(\mathbf{r}-\mathbf{r}^{\prime})$, after Fourier transformation to $\mathbf{k}$-space, is proportional to $\mathbf{k}/|\mathbf{k}|^2$ \cite{ruib,Ruia,Tigrana}. As has been shown in Ref.\cite{Ruia}, $ V^{(1)}_{int}$ will generally result in the inter-valley pairing of CS fermions, i.e., the CS superconductor state indicated by Fig.1(c). This is an alternative description of the N\'{e}el AFM ground state \cite{Ruia}.

Let us now turn on frustration by tuning $\mathbf{J}$ towards $\mathbf{J}_{fin}$. In this process, the dispersion away from the Dirac nodes $\mathbf{K}_a$ in Eq.\eqref{eq3} will be unavoidably perturbed. An expansion near $\mathbf{K}_a$ allows us to represent the low-energy action as $S_{XY}+\delta S$ \cite{Ruia}, where $\delta S$ consists of additional couplings between CS fermions and the CS gauge field $A_{\mu}$. Again, after integrating out $A_{\mu}$, a number of new interactions, $V^{(i)}_{int}$, $i=2,..M$, will be generated with different weights $w_i$, which is tunable by $\mathbf{J}$. Thus, for any interaction $\mathbf{J}$, the low-energy effective theory is given by
\begin{equation}\label{eq5}
  H_{f}=\sum^N_{a=1}\int d\mathbf{r}f^{\dagger}_{a,\mathbf{r}}\boldsymbol{\sigma}^{(a)}\cdot(-i\nabla)f_{a,\mathbf{r}} +\sum^M_{i=1}w_i(\mathbf{J})V^{(i)}_{int},
\end{equation}
which describes N Dirac CS fermions interacting via M  interaction terms. In general, the Pauli matrix $\boldsymbol{\sigma}^{(a)}$ can be valley-dependent such that the interacting Dirac fermions enjoys valley-dependent pseudospin-momentum locking (PSML) \cite{ruic}. $V^{(i)}_{int}$ relies on specific models, therefore not written explicitly here. 
We have shown that the original spin model with tunable frustration is exactly mapped in low-energy to a correlated Dirac CS fermions, $H_f$. The case for $N=2$ and $M=2$ is illuatrated by Fig.1(b).  It should be noted that Eq.\eqref{eq5} entirely captures the low-energy physics of the general 2D frustrated XY quantum magnets with tunable frustration, unifying all possible ground states, either confined or deconfined, in a fermionic language.

\emph{\color{blue}{CS excitonic insulator.--}}
We now apply the above framework to the $J_1$-$J_2$ XY honeycomb model. Following the procedure, we obtain the low-energy theory where Dirac CS fermions reside on two valleys and are subject to two interactions \cite{sup}, $H_{tot}=\sum_{a=\pm} H_{0,a}+V^{(1)}_{int}+V^{(2)}_{int}$, where
\begin{equation}\label{eqn7}
  H_{0,a}=\sum_{\mathbf{k}}f^{\dagger}_{a,\mathbf{k}}av_F\mathbf{k}\cdot\boldsymbol{\sigma}^{(a)}f_{a,\mathbf{k}}.
\end{equation}
$v_F=\sqrt{3}\epsilon J_1/2$, and $\epsilon$ is the lattice constant. A momentum cutoff $\Lambda$ is implicit. The two nodes are denoted by $a=\pm$ and $\boldsymbol{\sigma}^{(+)}=\boldsymbol{\sigma}$, $\boldsymbol{\sigma}^{(-)}=\boldsymbol{\sigma}^{\mathrm{T}}$, reflecting the PSML. The interaction, $V^{(1)}_{int}$, induced by nearest neighbor (n.n.) gauge field is derived as \cite{sup},
\begin{equation}\label{eq8}
\begin{split}
  V^{(1)}_{int}&=\sum_{\mathbf{k}_1,\mathbf{k}_2,\mathbf{q}}v^{\alpha\beta\rho\lambda}_{\mathbf{q}}f^{\dagger}_{a,\alpha,\mathbf{k}_1}f_{a,\beta,\mathbf{k}_1+\mathbf{q}}f^{\dagger}_{\overline{a},\rho,\mathbf{k}_2}f_{\overline{a},\lambda,\mathbf{k}_2-\mathbf{q}},
\end{split}
\end{equation} 
where $v^{\alpha\beta\rho\lambda}_{\mathbf{q}}=- e v_F(\sigma^i_{\alpha\lambda}\delta_{\beta\rho}+\delta_{\alpha\lambda}\sigma^{i T}_{\lambda\rho})A^j_{\mathbf{q}}$, with $A^j_{\mathbf{q}}=2\pi i \epsilon^{ij}q^j/q^2$, $\epsilon^{ij}$ being the Levi-Civita tensor.  $V^{(1)}_{int}$ is an intra-valley interaction of $p$-wave symmetry, dominant for weak frustration near $\mathbf{J}_{ini}$. The other interaction, generated by the n.n.n. gauge field,  is derived as \cite{sup},
\begin{equation}\label{eq9}
  V^{(2)}_{int}=\sum_{\mathbf{k}_1,\mathbf{k}_2,\mathbf{q}}\frac{\mathbf{k}_1
  \cdot\mathbf{A}_{\mathbf{q}}}{m}f^{\dagger}_{a,\alpha,\mathbf{k}_1}f_{a,\alpha,\mathbf{k}_1+\mathbf{q}}f^{\dagger}_{b,\beta,\mathbf{k}_2}f_{b,\beta,\mathbf{k}_2-\mathbf{q}},
\end{equation}
where $1/m=3eJ_2/2$.  We note that the frustration in the original spin model is now quantified by the competition between $V^{(1)}_{int}$ and $V^{(2)}_{int}$, which can be analyzed by many-body techniques for fermions.

Let us firstly focus on  $V^{(2)}_{int}$ only. It is irrelevant in perturbative renormalization group (RG) sense \cite{rshankar}, however, possible instabilities are still likely to occur for large interaction, $1/m$ \cite{ruie,hwei}. Besides, the vertex, $\mathbf{k}_1\cdot\mathbf{A}_{\mathbf{q}}$, is non-standard and generates new types of interactions in the perturbation expansion. Then, the leading instability should be predictable by studying the least irrelevant term in the generated interactions \cite{iaffleck}. Owing to the momentum angle dependence of the vertex, most diagrams vanish after integrating out the fast modes in one-loop order, required by the rotational invariance. The leading term shows up as \cite{sup},
\begin{equation}\label{eq10}
V^{(2)}_{eff}=v_{eff}\sum_{\mathbf{k}_1,\mathbf{k}_2,\mathbf{q}}f^{\dagger}_{a,\alpha,\mathbf{k}_1}f_{a,\alpha,\mathbf{k}_1+\mathbf{q}}f^{\dagger}_{b,\beta,\mathbf{k}_2}f_{b,\beta,\mathbf{k}_2-\mathbf{q}},
\end{equation}
where  $v_{eff}=2\pi^2/m^2\Lambda v_F$.  Therefore, apart from $V^{(1)}_{int}$, the effective Hamiltonian dominant in low-energy  is then given by $H_{eff}=\sum_{a=\pm} H_{0,a}+V^{(2)}_{eff}$.

Eq.\eqref{eq10} includes both the inter- and intra-valley couplings. Moreover, after transformation to the band basis via $f_{a,\pm,\mathbf{k}}=(\pm e^{-ia\theta}c_{a,+,\mathbf{k}}+c_{a,-,\mathbf{k}})/\sqrt{2}$ with $\theta$ being the angle of $\mathbf{k}$, Eq.\eqref{eq10} involves the inter-band scattering terms. For strong enough $v_{eff}$, such scattering is able to generate particle-hole instabilities, either forming an inter-valley charge density wave (CDW) \cite{hwei} or an intra-valley EI \cite{ruie}. Due to the PSML in $H_{0,a}$, both $s$-wave and $p$-wave particle-hole pairs are allowed by symmetry \cite{ruie}. We therefore decouple Eq.\eqref{eq10} by mean-field orders formed by CS fermions (CS orders), with either $s$- or $p$-wave symmetry. The intra-valley EI orders are respectively given by $\chi_s=v_{eff}\sum_{\mathbf{k}}\langle c^{\dagger}_{a,n,\mathbf{k}}c_{a,\overline{n},\mathbf{k}}\rangle$ and $\chi_p$ defined by $\frac{1}{\sqrt{2}}\chi_p(\hat{\mathbf{e}}_x+i\hat{\mathbf{e}}_y)\equiv v_{eff}\sum_{\mathbf{k}}(\mathbf{k}/|\mathbf{k}|) \langle c^{\dagger}_{a,n,\mathbf{k}}c_{a,\overline{n},\mathbf{k}}\rangle$. Similarly, CS orders  $\overline{\chi}_s$ and $\overline{\chi}_p$ are introduced for the inter-valley CDWs, which are particle-hole orders formed by CS fermions on different valleys.

\begin{figure}[tbp]\label{fig3}
\includegraphics[width=\linewidth]{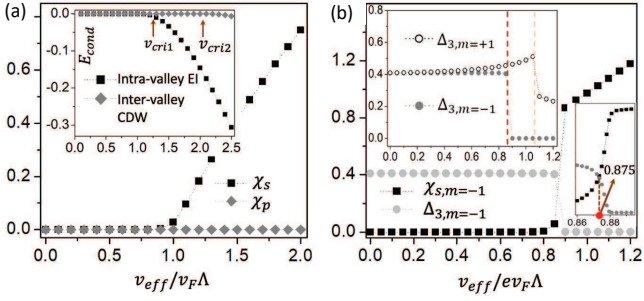}
\caption{(a) CS fermion mean-field solution for $H_{eff}$ (in unit of $v_F\Lambda$). A s-wave EI with nonzero $\chi_s$ is stabilized for $v_{eff}/v_F\Lambda>1$. Inset compares the condensation energy, $E_{cond}$, of the  $s$-wave intra-valley EI with that of the inter-valley CDW. The former has lower energy and smaller interaction threshold.  (b) The order parameters $\chi_s$ and $\Delta_3$ solved from $H_{tot}$ for the $\mathcal{T}$-breaking configuration ($m=-1$). Upper-left inset compares the $\Delta_3$ obtained for $m=-1$ and $m=+1$, which shows that the $\mathcal{T}$-breaking EI  ($m=-1$) is the leading instability of the CS superconductor \cite{Ruia}. }
\end{figure}

The self-consistent mean-field solution with respect to $H_{eff}$ is shown in Fig.3(a). One observes that, for large $v_{eff}$, the instability always occurs in the $s$-wave rather than the $p$-wave channel. As shown by the inset, $\chi_{s}$ emerges at the critical point $v_{cri1}=v_F\Lambda$, much earlier than $\overline{\chi}_s$, and it also has a lower condensation energy than the latter. Thus,  the intra-valley EI is the most stable CS order, if one considers the interaction $V^{(2)}_{int}$ alone.

It is worthwhile to note that the formation of the EI gaps out both the two Dirac nodes independently, with the mass gap $\chi_{a,s}$ as indicated by Fig.1(e). Thus, there are two energy degenerate configurations with either equal or opposite mass terms for the two valleys, i.e., $\chi_{a,s}=m\chi_{\overline{a},s}$ with $m=\pm1$.  Since  $\mathcal{T}$ is violated (respected) for $m=-1$ ($m=+1$),  the system does not tell whether $\mathcal{T}$ will be spontaneously broken or not. As will be clear below, this is only true when  $V^{(1)}_{int}$ is absent.



\emph{\color{blue}{Transition from CS superconductor to CS exciton insulator.--}}
We now turn on $V^{(1)}_{int}$, which favors a CS superconductor \cite{Ruia}.  Then, two types of CS order should be naturally considered as the ground state  candidates. One is the intra-valley EI, $\chi_s$, and the other is the inter-valley CS superconductor. The latter,  has order parameter in the 2 by 2 sublattice space \cite{Ruia}, $\hat{\Delta}_{\mathbf{k}}\equiv\Delta^{\rho\lambda}_{\mathbf{k}}=\sum_{\mathbf{k}^{\prime}}v^{\alpha\beta\rho\lambda}_{\mathbf{k}-\mathbf{k}^{\prime}}\langle f_{a,\alpha,\mathbf{k}^{\prime}}f_{\overline{a},\beta,-\mathbf{k}^{\prime}}\rangle$. Symmetry analysis further simplifies $\hat{\Delta}_{\mathbf{k}}$, leaving only one order parameter, i.e., the diagonal component $\Delta_3$, in the long-wave regime \cite{Tigrana,ruib}.  With these knowledge, we arrive at a global mean-field theory that describes the competition between the CS superconductor and the CS exciton insulator:
\begin{equation}\label{eq11}
  H^{(m)}_{tot}= v_F\mathbf{k}\cdot\boldsymbol{\sigma}\tau^3+\hat{\Delta}_{\mathbf{k}}\tau^++\hat{\Delta}^{\dagger}_{\mathbf{k}}\tau^-+\chi_s\tau^{(m)}\sigma^z,
\end{equation}
which is written in the Nambu-sublattice basis, $\Psi_{\mathbf{k}}=[f_{a,A,\mathbf{k}},f_{a,B,\mathbf{k}},f^{\dagger}_{\overline{a},A,-\mathbf{k}},f^{\dagger}_{\overline{a},B,-\mathbf{k}}]^\mathrm{T}$.  $m=-1$ ($+1$) denotes the aforementioned $\mathcal{T}$-breaking (-preserving) EI configuration, and  $\tau^{(\pm)}$ are defined as $\tau^{(+)}\equiv\tau^0$ and $\tau^{(-)}\equiv\tau^z$.

For both $m=\pm1$, we solve $\Delta_{3}$ and $\chi_s$ from the self-consistent equations derived from Eq.\eqref{eq11}. As shown in Fig.3(b), for $m=-1$, the CS superconductor is stable for small $v_{eff}$, displaying a full superconductor gap $\Delta_3$ indicated by Fig.1(c). With increasing $v_{eff}$ to around $v_{eff}=0.875ev_F\Lambda$, the CS superconductor is suppressed, accompanied by a fast growth of $\chi_s$. Further increasing $v_{eff}$, the system ends up into a stable CS exciton insulator (Fig.1(e)).  The upper-left inset to Fig.3(b) makes comparison between the $m=+1$ and $m=-1$ configuration. It is clearly found that the $\mathcal{T}$-breaking EI takes place as the leading instability of the CS superconductor \cite{Ruia} at a much earlier transition point. Recalling that $m=\pm1$ are degenerate when only $V^{(2)}_{int}$ is considered, the breaking of  $\mathcal{T}$ here is therefore a result of the interplay between $V^{(1)}_{int}$ and $V^{(2)}_{int}$, and thus the competition between the CS superconductor and exciton insular. The right inset identifies the critical point to be $v_{eff}/ev_F\Lambda=0.875$. Further taking into account $v_{eff}=2\pi^2/m^2\Lambda v_F$, $\Lambda=\pi$, and the smallest physical CS charge $e=3$ \cite{Ruia,Tigrana,ruib}, we obtain that the earliest transition takes place at $J_2/J_1\simeq0.22$.

\emph{\color{blue}{Non-uniform CSL and effective field theory}}.--
We now examine more closely on the CS exciton insulator. For $v_{eff}/ev_F\Lambda \gg 0.875$, $\Delta_3=0$, it is described by,
\begin{equation}\label{eq12}
  H_{CS-EI}=v_F(k_x\tau^z\sigma^x+k_y\tau^0\sigma^y)+\chi_s\tau^z\sigma^z.
\end{equation}
Here, the second term introduces a staggered chemical potential for CS fermions on A and B sublattice. This brings about the out-of-plane N\'{e}el ordering in the spin language. Furthermore, it also breaks $\mathcal{T}$ and generates nontrivial band topology with Chern number $C=1$, implying the chiral edge state, a reminiscence of the Haldane's model on honeycomb lattice \cite{qah}.

So far, we have neglected the fluctuation of the order parameter. The fluctuation behaves as a $\mathrm{U}(1)$ gauge field $A_{\mu}$ attached to $\chi_s$ \cite{xgwena}.  Due to the nontrivial band topology,  the gauge fluctuation receives nontrivial renormalization and manifests a CS gauge term, $S_{CS}=(1/4\pi)\int d^3r\epsilon^{\mu\nu\rho}A_{\mu}\partial_{\nu}A_{\rho}$, as a result of the parity anomaly of the 2+1D massive Dirac fermions \cite{redlich}. Furthermore, since $V^{(1)}_{int}$ is no longer relevant for $v_{eff}/ev_F\Lambda \gg 0.875$, it should be written back to the form as in Eq.\eqref{eq3}, generating another CS term. Collecting both terms, we arrive at the gauge theory,
$S_{CS-EI}=\frac{K}{4\pi}\int d^3r\epsilon^{\mu\nu\rho}A_{\mu}\partial_{\mu}A_{\rho}$ where $K=2$. The CS term stabilizes the fractionalized mean-field state in Eq.\eqref{eq12}, as it gaps out the gauge field $A_{\mu}$ \cite{xgwena}. Moreover,  $K=2$ is a smoking-gun indication of a CSL with semionic excitations \cite{vkalmeyerr,xgwena}. Therefore, the above results claim a UPT from the N\'{e}el order to a non-uniform CSL, in mutual support with our tensor-network-based calculations.
 
Now we highlight the significance of Eq.(9), a key result of this work. Although it appears in a simple mean-field
form, it elegantly describes the UPTs in the frustrated spin model and provides information about the phases separated by the UPT. To obtain Eq.(9), we first constructed a fermionized Hamiltonian that unites the UPT and adjacent phases into a single physical picture. Within this picture, the Neel AFM and the fractionalized states are both described as certain CS orders formed by the CS fermions. Secondly, we carefully investigated the interactions between CS fermions and systematically selected the most stable CS order candidates, which were not apparent in the original spin language. The competition between the CS orders is fully captured by Eq.(9), which therefore provides a concise characterization of the UPT.


\emph{\color{blue}{Conclusion and discussion.--}}
We propose a framework to investigate the presence of UPTs in 2D frustrated XY quantum magnets, based on the study of competition between different CS orders. The proposed method is lattice-independent, and as such, it is general and applicable to a variety of frustrated 2D systems. For the spin-1/2 $J_1$ -$J_2$  XY model on the honeycomb lattice, the method predicts a phase transition from the 
CS superconductor to the CS exciton insulator. In the language of original spin operators, this transition  essentially describes a UPT from the Neel AFM to a non-uniform CSL with U(1) gauge fluctuations. Our high-precision tensor network calculation confirms these analytical findings and provides the first supportive evidence for a non-uniform CSL. Furthermore, our method is generalizable to XXZ and Heisenberg models, which are more relevant to realistic materials.  Also, the proposed concept of CS order is not exclusive to Neel AFM and CSL but can be generalized to classify and describe more types of QSLs.

The proposed approach enables a systematic study of the competing CS orders, which in turn can provide the effective field theory of the emergent criticality. Therefore, the reported work may have broader applications and promise a general recipe for the studies of UPTs in a variety of strongly-correlated many-body systems.



%
\begin{acknowledgments}
Z. Y. X. acknowledges Hong-Hao Tu for helpful discussions. This work was supported by the Youth Program of National Natural Science Foundation of China (Grant No. 11904225 and 11774420) and  the National Key  R\&D Program of China (Grant No. 2017YFA0303200, 2016YFA0300503, and 2017YFA0302900). T.A.S. acknowledges startup funds from UMass Amherst.
\end{acknowledgments}

\pagebreak
\vspace{5cm}
\widetext
\setcounter{equation}{0}
\setcounter{figure}{0}
\setcounter{table}{0}
\setcounter{page}{1}
\makeatletter
\renewcommand{\theequation}{S\arabic{equation}}
\renewcommand{\thefigure}{S\arabic{figure}}
\renewcommand{\bibnumfmt}[1]{[S#1]}
\renewcommand{\citenumfont}[1]{S#1}

\pagebreak
\vspace{5cm}
\widetext
\begin{center}
\textbf{\large Supplemental material for: Theory of competing Chern-Simons orders and emergent phase transitions}
\end{center}

\section{Tensor network study of the perturbed $J_1$-$J_2$ XY model}
The frustrated XY model on honeycomb lattice we studied in this work is defined in the following
\begin{equation}
H = J_1\sum_{\langle i,j\rangle}(S^{+}_iS^{-}_{j} + h.c.) + J_2\sum_{\langle\langle i,j\rangle\rangle}(S^{+}_iS^{-}_{j} + h.c.)
\end{equation}
In order to study the possible chiral phase more clearly, we introduced a flux as weak perturbation in each triangle but with opposite sign, that is we associated each upper-triangle with the following extra term
\begin{equation}
H^{(abc)}_p = \varepsilon\cdot\left(e^{i\theta_{ab}}S^{+}_{a}S^{-}_{b} + e^{i\theta_{bc}}S^{+}_{b}S^{-}_{c} + e^{i\theta_{ca}}S^{+}_{c}S^{-}_{a} + h.c.\right)
\end{equation}
where $(a,b,c)$ are the site indices arranged clockwisely associated with the upper triangle, and similarly there is a $H^{(abc)}_p$ term introduced for each lower-triangle but the site indices $(a,b,c)$ are arranged counter-clockwisely. In our calculation, we set
\begin{equation}
\varepsilon = 10^{-4}, \quad \theta_{ab} = \theta_{bc} = \theta_{ca} = \frac{\pi}{2}
\end{equation}
for all the triangles. In addition, in order to stabilize the intermediate phase and speed up the convergence, we introduced also a very small staggered magnetic field in z-direction with magnitude $10^{-8}$ at each lattice site.

\begin{figure}[htb]
\includegraphics[width=0.5\linewidth]{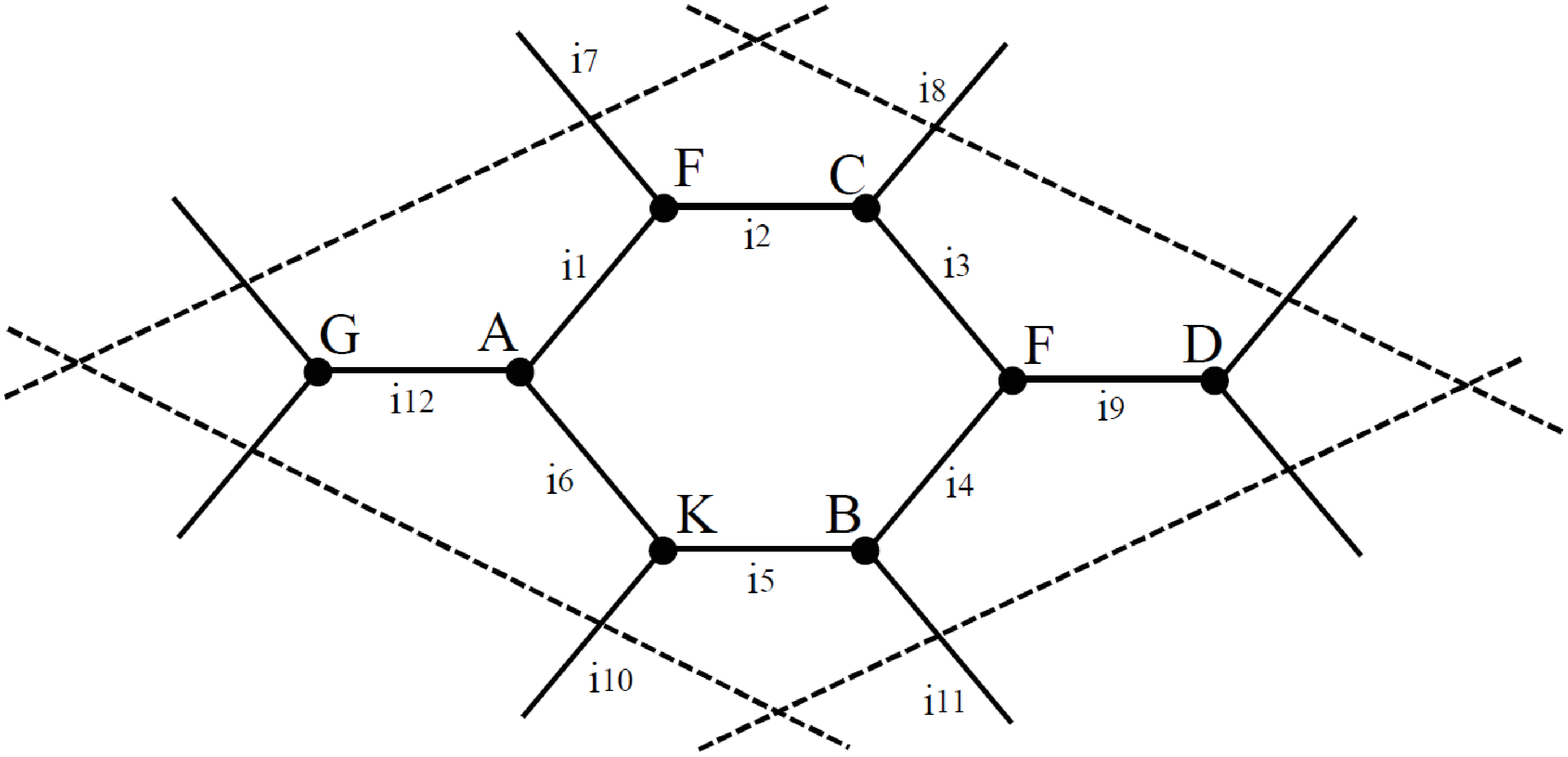}
\caption{Illustration of the unit cell used in the PESS wavefunction ansatz. The spin indices $\{\sigma\}$ defined on lattice sites appeared in Eq.~(\ref{Eq:WF}) are not shown explicitly.}
\label{Fig:WF}
\end{figure}

The ground state is represented as a tensor-network ansatz, more precisely the projected entangled simplex state (PESS) \cite{PESS2014}, defined in the following
\begin{equation}
|\Psi\rangle = \sum_{\{\sigma\}}\sum_{\{i\}}\prod_{\alpha}A^{\alpha}_{i_1i_{12}i_6}[\sigma_a]B^{\alpha}_{i_4i_5i_{11}}[\sigma_b]C^{\alpha}_{i_8i_2i_3}[\sigma_c]
    D^{\alpha}_{i_{10}i_9i_7}[\sigma_d]E^{\alpha}_{i_1i_2i_7}[\sigma_e]F^{\alpha}_{i_4i_9i_3}[\sigma_f]G^{\alpha}_{i_8i_{12}i_{11}}[\sigma_g]
    K^{\alpha}_{i_{10}i_5i_6}[\sigma_h]|\sigma\rangle
\label{Eq:WF}
\end{equation}
where $\alpha$ is the index of unit cell, $i$ is the bond index, and $\sigma$ is the spin index. As shown in Eq.~(\ref{Eq:WF}) and illustrated in Fig.~(\ref{Fig:WF}), we choose a unit cell consisting of eight different tensors in order to distinguish the three phases.
To determine the parameters contained in the wave function, we use the imaginary-time evolution method applied to an arbitrary initial state $|\Psi_0\rangle$, in which a small trotter step is decomposed into eight substeps (due to the eight sublattice structure) through Trotter-Suzuki decomposition. E.g., the evolution substep applied to the small triangle centered as site A has the form $e^{-\tau H_{A}}$, where $\tau$ is the trotter step and the local Hamiltonian $H_A$ can be written as
\begin{eqnarray}
H_{A} = \frac{J_1}{2}(S^{+}_{G}S^{-}_{A} + S^{+}_{F}S^{-}_{A} + S^{+}_{K}S^{-}_{A})
    + J_2(S^{+}_{G}S^{-}_{F} + S^{+}_{F}S^{-}_{K} + S^{+}_{K}S^{-}_{G}) + h.c.
\label{Eq:LocalH}
\end{eqnarray}
The factor $1/2$ comes from double-counting, and the subscripts of the spin operators denote the site indices to which the tensors are associated. The perturbation $H_p$ and the staggered field can be added to Eq.~(\ref{Eq:LocalH}) easily.

Using the simple update algorithm \cite{SU1D2007, SU2D2008}, we follow the standard evolution procedure for PESS wave function elaborated in Ref.~\cite{PESS2014}. In order to further refine the state, we perform no more than 50 steps of full update \cite{FU2014} after the representation is converged. In the determination of the environment used in both full update and observable calculation, we deform the honeycomb lattice into square lattice and employed the corner transfer-matrix renormalization group method \cite{CTMRG1996, CTMRG2009, CTMRG2014} for $2\times 2$ sublattice structure, in which the boundary dimension $\chi$ is kept no more than 120 so that the calculation can be performed effectively.

\begin{figure}[htb]
\includegraphics[width=0.5\linewidth]{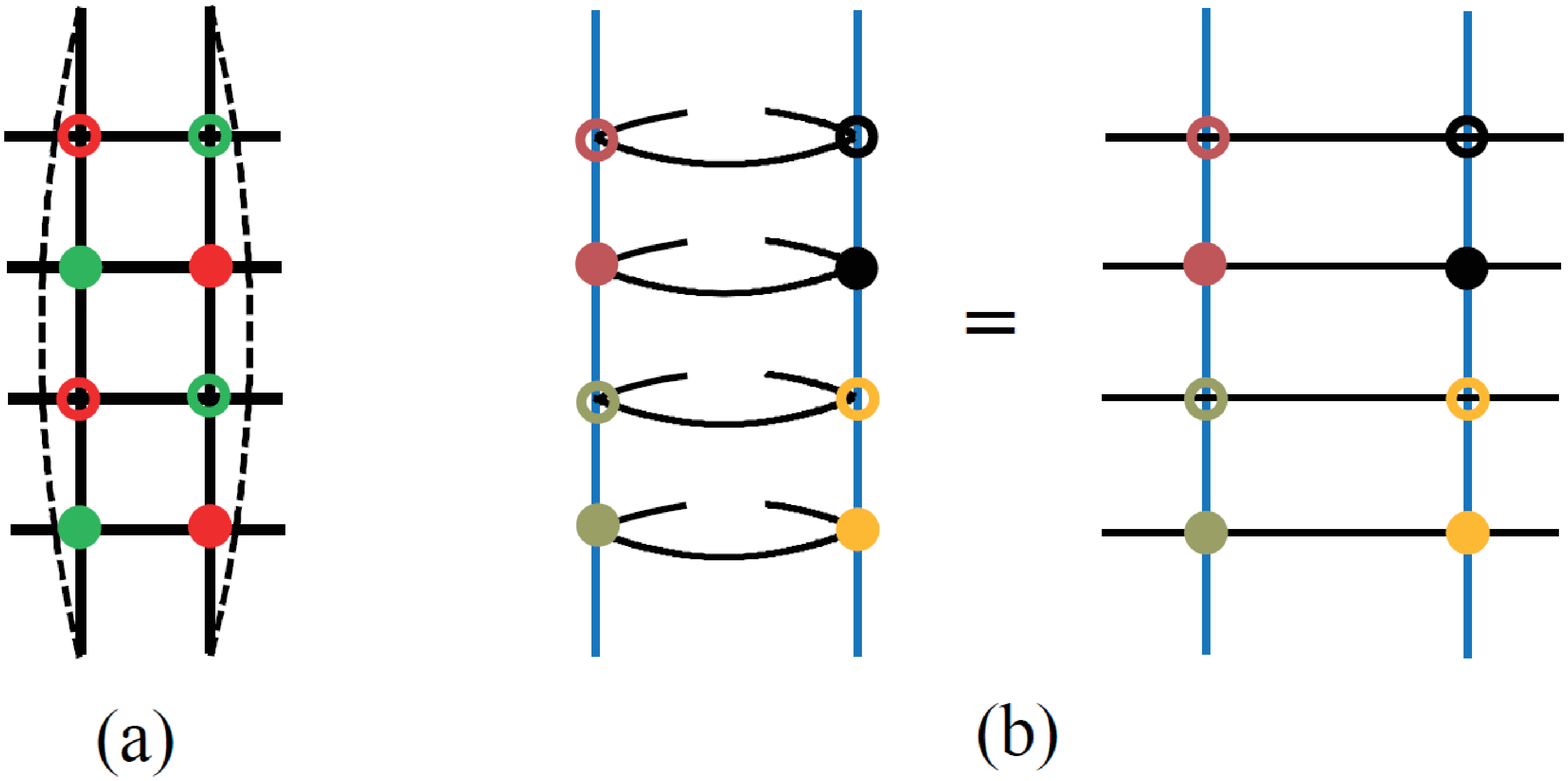}
\caption{Illustration of entanglement spectra calculation for cylinder with $L_y = 2$. (a) The transfer matrix represented as MPO. The four different markers denote different reduced tensors from the original tensor in the square lattice obtained by deformation of honeycomb. The bold lines denote squared bond dimensions. Dashed lines denote PBC. (b) The dominant eigenvector $\sigma_L$ and $\sigma_R$ of the transfer matrix shown in (a), represented as MPS, and their product $\sigma_L^{T}\sigma_R$ represented as MPO. }
\label{Fig:ESlattice}
\end{figure}

To check the chiral nature, we place the obtained wave function, deformed to square lattice, in an infinite long cylinder with circumference $L_y$ no more than 6 (unit cells) in the direction with periodic boundary condition (PBC), and then calculate the entanglement spectra associate with the bipartition of the cylinder in the direction with open boundary condition (OBC). Following the procedure elaborated in Ref.~\cite{Cirac2011, Poil2016, Saeed2018}, we firstly calculate the dominant eigenvectors, $\sigma_L$ and $\sigma_R$ as matrix product states (MPS), of the related transfer matrix constructed by the reduced tensors in the direction of PBC, and then represent $\sigma_L^{T}\sigma_R$ as the form of matrix product operator (MPO), whose eigenvalues gives the required entanglement spectra $\lambda$'s. This procedure is sketched in Fig.~(\ref{Fig:ESlattice}).

It is useful to calculate the entanglement spectra in each lattice momentum $k$-sector, with $k$ is the good quantum number associated with the translational invariance of the MPO. In our work, we construct the Krylov-subspace for a given $k$, and determine the first $q$ largest eigenvalues in each $k$-space, where $q$ is no more than 35. To be specific, we use the following ansatz to characterize a vector in the Krylov-space with momentum $k$,
\begin{equation}
|\Phi\rangle = \sum_{m=0}^{n-1}e^{imk}\hat{T}^{m}|\phi\rangle
\label{Eq:TIMPS}
\end{equation}
where $\hat{T}$ is the translation operator satisfying the following equation
\begin{equation}
\hat{T}|\Phi\rangle = e^{-ik}|\Phi\rangle, \quad \text{wth} \quad k = 0, \frac{2\pi}{n}, ..., \frac{2\pi(n-1)}{n}
\end{equation}
which can be verified easily. In Eq.~(\ref{Eq:TIMPS}), $|\phi\rangle$ is an arbitrary vector, and in our calculations, we represent $\phi$, and thus each vector generated in the Arnoldi iterations, as MPS. The most important thing we need to determine is the overlap of those arnoldi basis, which can be done in the situation of finite MPS. The details here are very similar with the Lanczos-MPS method we developed before to study the ground state and lowest excitation in Ref.~\cite{LanczosMPS}.

In our calculation, actually we calculate part of spectra $\{\lambda\}$, from which we determine the entanglement entropy (after proper normalization) $S$. By fitting the scaling behavior \cite{Kitaev2006, Wen2006, Eisert2010}
\begin{equation}
S \sim \alpha L_{y} - \gamma + O(1/L_y)
\end{equation}
with $L_y$ ranging from 3 to 6, we obtain a non-zero topological entanglement entropy $\gamma$ about 0.11, implying that a topological ordered state is obtained, whereas, it is smaller than the predicted value $(\ln{2})/2$, possibly due to the incompleteness of entanglement spectra.

\section{derivation of the interacting fermion model on honeycomb lattice}
We start from the $J_1$-$J_2$ XY model on honeycomb lattice
\begin{equation}\label{eqn1}
  H=J_1\sum_{\mathbf{r},j}\hat{S}^{+}_{\mathbf{r}}\hat{S}^-_{\mathbf{r}+\mathbf{e}_j}
  +J_2\sum_{\mathbf{r},j}\hat{S}^{+}_{\mathbf{r}}\hat{S}^-_{\mathbf{r}+\boldsymbol{\mu}_j}+h.c.,
\end{equation}
where a global factor $1/2$ has been absorbed. $\hat{S}^{\pm}_{\mathbf{r}}$ are the spin raising/lowering operators. We use $\mathbf{e}_j$ and $\boldsymbol{\mu}_j$ to denote the lattice vectors between the nearest and next nearest neighbour (n.n. and n.n.n.) bonds. Upon CS fermionization, the single-particle sector can be obtained as following with ``turning off" the CS gauge field,
\begin{equation}\label{eqn2}
  H_0= J_1\sum_{\mathbf{r},j}f^{\dagger}_{\mathbf{r}}f_{\mathbf{r}+\mathbf{e}_j}
  +J_2\sum_{\mathbf{r},j}f^{\dagger}_{\mathbf{r}}f^-_{\mathbf{r}+\boldsymbol{\mu}_j}+h.c.
\end{equation}
Making transformation to $\mathbf{k}$-space and expansion near two valleys denoted by $\pm$, (i.e., the two valleys of the honeycomb lattice), the total Hamiltonian is cast in low energy to a sum of Dirac Hamiltonians with $\mathbf{k}^2$ corrections (due to the n.n.n hopping) near the two valleys, $H_0=H_{0,+}+H_{0,-}$, where the Hamiltonian at $a=\pm$ valley reads as,
\begin{equation}\label{eqn3}
  H_{0,a}=\sum_{\mathbf{r}}f^{\dagger}_{a,\mathbf{r}}[\frac{\hat{\mathbf{k}}^2}{2m^{\prime}}\sigma^0\pm v_F\hat{\mathbf{k}}\cdot\boldsymbol{\sigma}^{(T)}]f_{a,\mathbf{r}},
\end{equation}
where $\hat{\mathbf{k}}=-i\boldsymbol{\nabla}$ is the momentum operator, $\boldsymbol{\sigma}^{(T)}$, the Pauli matrix denoting the sublattice space, is defined to be $\boldsymbol{\sigma}$ for $K$ valley and the transposition $\boldsymbol{\sigma}^T$ for $K^{\prime}$ valley, $m^{\prime}=2/(3J_2)$, and $v_F=\sqrt{3} J_1/2$. Eq.\eqref{eqn3} is the single-particle Hamiltonian in terms of CS fermions, which can serve as a  free fixed point in RG sense that may develop instabilities under additional gauge-induced interactions.

With turning on the CS gauge field, the total Hamiltonian formally reads as,
\begin{equation}\label{eqn4}
  H=J_1\sum_{\mathbf{r},j}f^{\dagger}_{\mathbf{r}}f_{\mathbf{r}+\mathbf{e}_j}e^{ie\mathcal{A}_{\mathbf{r},\mathbf{r}+\boldsymbol{e}_j}}
  +J_2\sum_{\mathbf{r},j}f^{\dagger}_{\mathbf{r}}f_{\mathbf{r}+\boldsymbol{\mu}_j}e^{ie\mathcal{A}_{\mathbf{r},\mathbf{r}+\mu_j}}+h.c.
\end{equation}
where $\mathcal{A}_{\mathbf{r},\mathbf{r}^{\prime}}=\sum_{\tilde{\mathbf{r}}}[\mathrm{arg}(\mathbf{r}-\tilde{\mathbf{r}})
-\mathrm{arg}(\mathbf{r}^{\prime}-\tilde{\mathbf{r}})]$ and $e$ the CS charge. The presence of the gauge field in Eq.\eqref{eqn4} can be absorbed via a $\mathrm{U}(1)$ gauge transformation, leading to $\hat{\mathbf{k}}\rightarrow\hat{\mathbf{k}}+\mathcal{A}_{\mathbf{r}}$, where $\mathcal{A}_{\mathbf{r}}$ is the continuum form of the lattice gauge field \cite{Ruia}. In low-energy regime, it can be derived from the string operator from the CS transformation as, $\mathcal{A}_{\mathbf{r}}=\sum_{\mathbf{r}^{\prime}}\mathbf{A}_{\mathbf{r}-\mathbf{r}^{\prime}}\sum_{a=\pm}f^{\dagger}_{a,\mathbf{r}^{\prime}}f_{a,\mathbf{r}^{\prime}}$,
where $A^i_{\mathbf{r}-\mathbf{r}^{\prime}}=\epsilon^{ij}(\mathbf{r}-\mathbf{r}^{\prime})_j/|\mathbf{r}-\mathbf{r}^{\prime}|^2$, with $\epsilon^{ij}$ being the Levi-Civita  rank-2 antisymmetric tensor. After the the gauge transformation, the total low-energy Hamiltonian is derived as
\begin{equation}\label{eqn5}
  H=\sum_{a}\sum_{\mathbf{r}}f^{\dagger}_{a,\mathbf{r}}[\frac{1}{2m^{\prime}}\sigma^0(\hat{\mathbf{k}}+e\sum_{\mathbf{r}^{\prime}}
  \mathbf{A}_{\mathbf{r}-\mathbf{r}^{\prime}}\sum_{b}f^{\dagger}_{b,\mathbf{r}^{\prime}}f_{b,\mathbf{r}^{\prime}})^2]f_{a,\mathbf{r}}+\sum_{a}\sum_{\mathbf{r}}f^{\dagger}_{a,\mathbf{r}}a v_F(\hat{\mathbf{k}}+e\sum_{\mathbf{r}^{\prime}}\mathbf{A}_{\mathbf{r}-\mathbf{r}^{\prime}}\sum_{b}f^{\dagger}_{b,\mathbf{r}^{\prime}}f_{b,\mathbf{r}^{\prime}})\cdot\boldsymbol{\sigma}f_{a,\mathbf{r}},
\end{equation}
where $f_{a,\mathbf{r}}=[f_{a,A,\mathbf{r}},f_{a,B,\mathbf{r}}]^T$ is a spinor in sublattice space. We see that the second term in Eq.\eqref{eqn5}, which comes from the original n.n. interaction, leads to a single nonlocal interaction between CS fermions,
\begin{equation}\label{eqn6}
  V^{(1)}_{int}=e\sum_{\mathbf{r},\mathbf{r}^{\prime}a}a v_F\mathbf{A}_{\mathbf{r}-\mathbf{r}^{\prime}}\cdot\boldsymbol{\sigma}f^{\dagger}_{a,\mathbf{r}}f_{a,\mathbf{r}}f^{\dagger}_{\overline{a},\mathbf{r}^{\prime}}f_{\overline{a},\mathbf{r}^{\prime}}.
\end{equation}
where one only needs to keep the inter-valley interaction while neglect the intra-valley one because of the priori knowledge on the N\'{e}el AFM state which requires a condensation at $\Gamma$ point. Eq.\eqref{eqn6} is transformed to the interaction Eq.(7) in the main text, after further inserting $\mathbf{A}_{\mathbf{r}-\mathbf{r}^{\prime}}$ and making transformation to momentum space.

The first term Eq.\eqref{eqn5}, which comes from the n.n.n exchange term $J_2$, contributes additional interactions:
\begin{equation}\label{eqn7}
  V^{(2)}_{int}=\frac{1}{m}\sum_{\mathbf{r},\mathbf{r}^{\prime}}\sum_{\rho,\lambda}\hat{\mathbf{k}}
  \cdot\mathbf{A}_{\mathbf{r}-\mathbf{r}^{\prime}}f^{\dagger}_{a,\mathbf{r}}f_{a,\mathbf{r}}f^{\dagger}_{b,\mathbf{r}^{\prime}}f_{b,\mathbf{r}^{\prime}},
\end{equation}
and a higher order term,
\begin{equation}\label{eqn8}
  V^{(3)}_{int}=\frac{1}{2m}\sum_{\mathbf{r},\mathbf{r}^{\prime},\mathbf{r}^{\prime\prime}}\sum_{a,b,c}\mathbf{A}_{\mathbf{r}-\mathbf{r}^{\prime}}
  \cdot\mathbf{A}_{\mathbf{r}-\mathbf{r}^{\prime\prime}}f^{\dagger}_{a,\mathbf{r}}f_{a,\mathbf{r}}f^{\dagger}_{b,\mathbf{r}^{\prime}}f_{b,\mathbf{r}^{\prime}}f^{\dagger}_{c,\mathbf{r}^{\prime\prime}}f_{c,\mathbf{r}^{\prime\prime}},
\end{equation}
where $m=m^{\prime}/e$. $V^{(3)}_{int}$ is a higher order correction with respect to the emergent gauge field. We expect that the critical behavior with respect to the first UPT taking place with increasing $J_2/J_1$ should be attributed to the lower order correction $V^{(2)}_{int}$, therefore the effect of $V^{(3)}_{int}$ is not considered in this work but could be an interesting topic for further studies. In momentum space, $V^{(2)}_{int}$ is cast into,
\begin{equation}\label{eqn9}
  V^{(2)}_{int}=\frac{1}{m}\sum_{\mathbf{k}_1,\mathbf{k}_2,\mathbf{q}}\mathbf{k}_1
  \cdot\mathbf{A}_{\mathbf{q}}f^{\dagger}_{a,\mathbf{k}_1}f_{a,\mathbf{k}_1+\mathbf{q}}f^{\dagger}_{b,\mathbf{k}_2}f_{b,\mathbf{k}_2-\mathbf{q}},
\end{equation}
where $\mathbf{A}_{\mathbf{q}}=2\pi i\epsilon^{ij}q^i/q^2$, thus we have derived Eq.(6),(7) of the main text after writing the sublattice indices explicitly.

On the other hand, before one integrates out the gauge field, a low-energy effective Hamiltonian can be derived from Eq.\eqref{eqn4}. The gauge field minimally couples to the two Dirac cones, generating

\begin{equation}\label{eqnew9}
  S_{XY}=\sum_a\int d\mathbf{r}dtf^{\dagger}_{a,\alpha,\mathbf{r}}(\mathbf{K}_i)\sigma^0_{\alpha\beta}(i\partial_0-A^0_{\mathbf{r}})f_{a, \mathbf{r},\beta}
  -\sum_a\int d\mathbf{r}dtf^{\dagger}_{a,\mathbf{r},\alpha}\sigma^i_{\alpha\beta}(-i\partial_i+eA_{\mathbf{r}})f_{a,\mathbf{r},\beta}+S_{CS},
\end{equation}
which is further written more compactly as the Eq.(2) of the main text.

\section{the perturbative renormalization group analysis on the interacting Dirac CS fermions}
\begin{figure}[htb]
\includegraphics[width=0.8\linewidth]{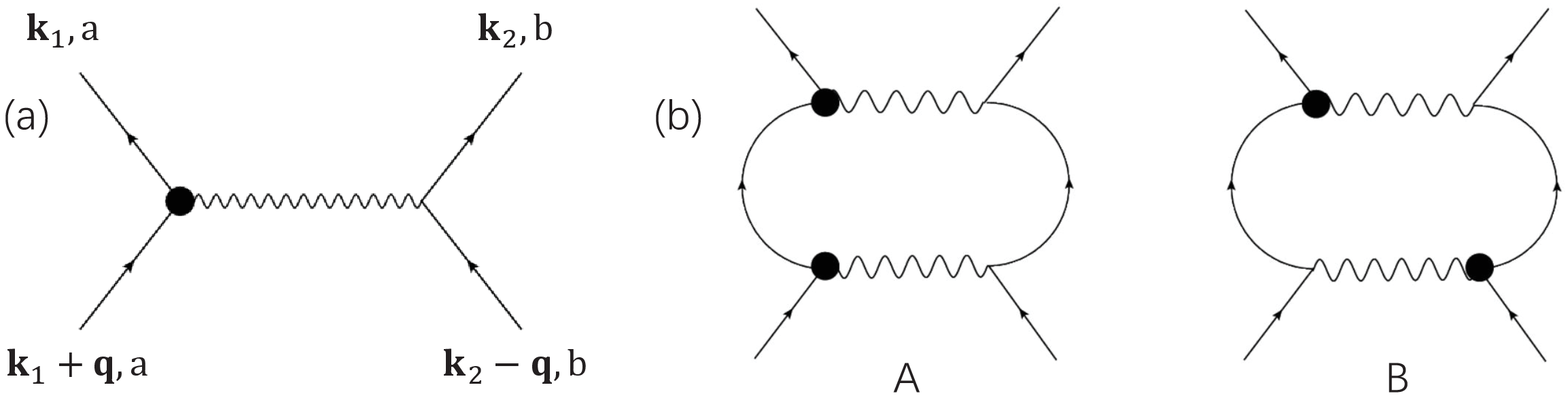}
\caption{(color online). (a) The interaction vertex of $V^{(2)}_{int}$ between CS fermions. (b) The BCS diagrams in second order renormalization.}
\label{figs1}
\end{figure}

\begin{figure}[htb]
\includegraphics[width=0.7\linewidth]{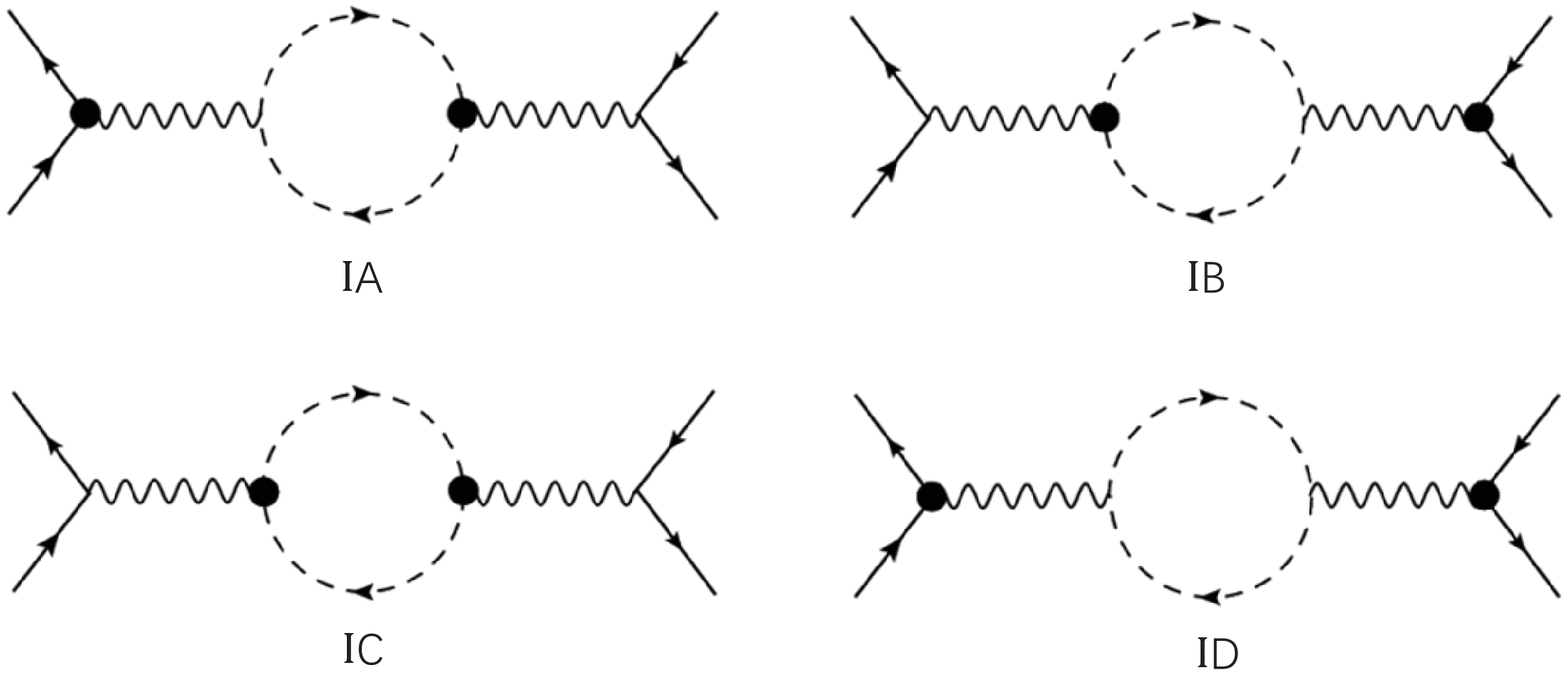}
\caption{(color online). class-I of the particle-particle diagrams in second order renormalization, where the internal propagators are dashed, denoting the integral of their internal momentums are restricted to the fast mode shell.}
\label{figs2}
\end{figure}

In this section, we study the effect of $V^{(2)}_{int}$ on the free CS fermions $H_0$. Since we are interested in small $k$, the $k^2$ correction in $H_0$ can be neglected in the long-wave length regime, leading to the action in the Euclidean space time, which describes free CS Dirac fermions on two valleys $a=\pm$,
\begin{equation}\label{eqn10}
  S_0=\sum_{a=\pm}\int\frac{d\omega}{2\pi}\int\frac{d^2k}{(2\pi)^2}f^{\dagger}_{a,\mathbf{k}}
  (i\omega-a v_F\boldsymbol{\sigma}\cdot\mathbf{k})f_{a,\mathbf{k}},
\end{equation}
The action describing the interaction $V^{(2)}_{int}$ then reads as,
\begin{equation}\label{eqn11}
  S^{(2)}_{int}=i\frac{2\pi}{m}\int\frac{d\omega_1dk^2_1}{(2\pi)^3}\int\frac{d\omega_2dk^2_2}{(2\pi)^3}\int\frac{d\nu dq^2}{(2\pi)^3}\frac{\mathbf{k}_1\times\mathbf{q}}{q^2}f^{\dagger}_{a,k_1}f_{a,k_1+q}f^{\dagger}_{b,k_2}f_{b,k_2-q}.
\end{equation}
After rescaling, it is clear that the scaling dimension of $S^{(2)}_{int}$ is -1, indicating that $V^{(2)}_{int}$ is an irrelevant perturbation for the CS Dirac fermions. To simplify notations, we introduce $\int_k=\int d\omega d^2k/(2\pi)^3$ and use $k$ to represent for the 3-vector $(\omega,\mathbf{k})$ in what follows, then we rewrite $S^{(2)}_{int}$ as,
\begin{equation}\label{eqn12}
  S^{(2)}_{int}=\int_{k_1,k_2,q}v^{ab}_{\alpha\alpha^{\prime}\beta\beta^{\prime}}(\mathbf{k}_1,\mathbf{q})f^{\dagger}_{a,\alpha,k_1}f_{a,\alpha^{\prime},k_1+q}f^{\dagger}_{b,\beta,k_2}f_{b,\beta^{\prime},k_2-q}.
\end{equation}
with the vertex dependent on valley and sublattice degrees of freedom as,
\begin{equation}\label{eqn13}
  v^{ab}_{\alpha\alpha^{\prime}\beta\beta^{\prime}}(\mathbf{k}_1,\mathbf{q})=iv_2\frac{\mathbf{k}_1\times\mathbf{q}}{q^2}\delta_{\alpha,\alpha^{\prime}}\delta_{\beta,\beta^{\prime}},
\end{equation}
where $v_2=2\pi/m$. The action $S^{(2)}_{int}$ can be represented by Feynman diagrams, as shown in Fig.\ref{figs1}. Since the vertex being proportional to $(\mathbf{k}_1\times\mathbf{q})/q^2$ in $S^{(2)}_{int}$, we use a shaded circle to denote this term in the diagram, which appear asymmetrically in the vertex as it is only dependent on momentum $\mathbf{k}_1$ rather than $\mathbf{k}_2$. This particular structure, derived as an intrinsic feature from the frustrated exchange couplings, will play important role and result in interesting possible instabilities of the free CS fermions, as will be shown below. The external legs in Fig.\ref{figs1} represent for the propagator of the CS fermions, a two by two matrix in the sublattice space. From $S_0$, the propagator is obtained for valley $a=\pm$ as,
\begin{equation}\label{eqn13}
  G^{(a)}(k)=-\frac{1}{\omega^2+k^2v^2_F}\left(
                                                           \begin{array}{cc}
                                                             i\omega & a v_F(k_x-ia k_y) \\
                                                             a v_F(k_x+ia k_y) & i\omega \\
                                                           \end{array}
                                                         \right).
\end{equation}
We consider the renormalization to one-loop order, which can be generally classified into BCS diagrams where particle-particle excitations take place in the internal loop, and the particle-hole diagrams.  Two inequivalent BCS diagrams are shown by Fig.\ref{figs1}(b). A specific calculation then directly shows that the diagram A and diagram B cancel with each other, due to the intrinsic asymmetric structure of $V^{(2)}_{int}$. This indicates that $V^{(2)}_{int}$, in contrast with $V^{(1)}_{int}$ that favors CS superconductors, is unlikely to induce any instability towards pairings of CS fermions. We then calculate the particle-hole diagrams, which we group into three classes I, II, and III according to their diagrammatic geometry. The diagrams in I-class are shown in Fig.\ref{figs2}.
A straightforward calculation directly reveals that the diagrams IA,B do not make any finite contributions due to the requirement of rotational invariance. The diagrams IC and ID also vanish due to the cancellation between $G^{(\rho)}_{11}G^{(\rho)}_{11}$ and $G^{(\rho)}_{12}G^{(\rho)}_{21}$. 
\begin{figure}[tb]
\includegraphics[width=0.7\linewidth]{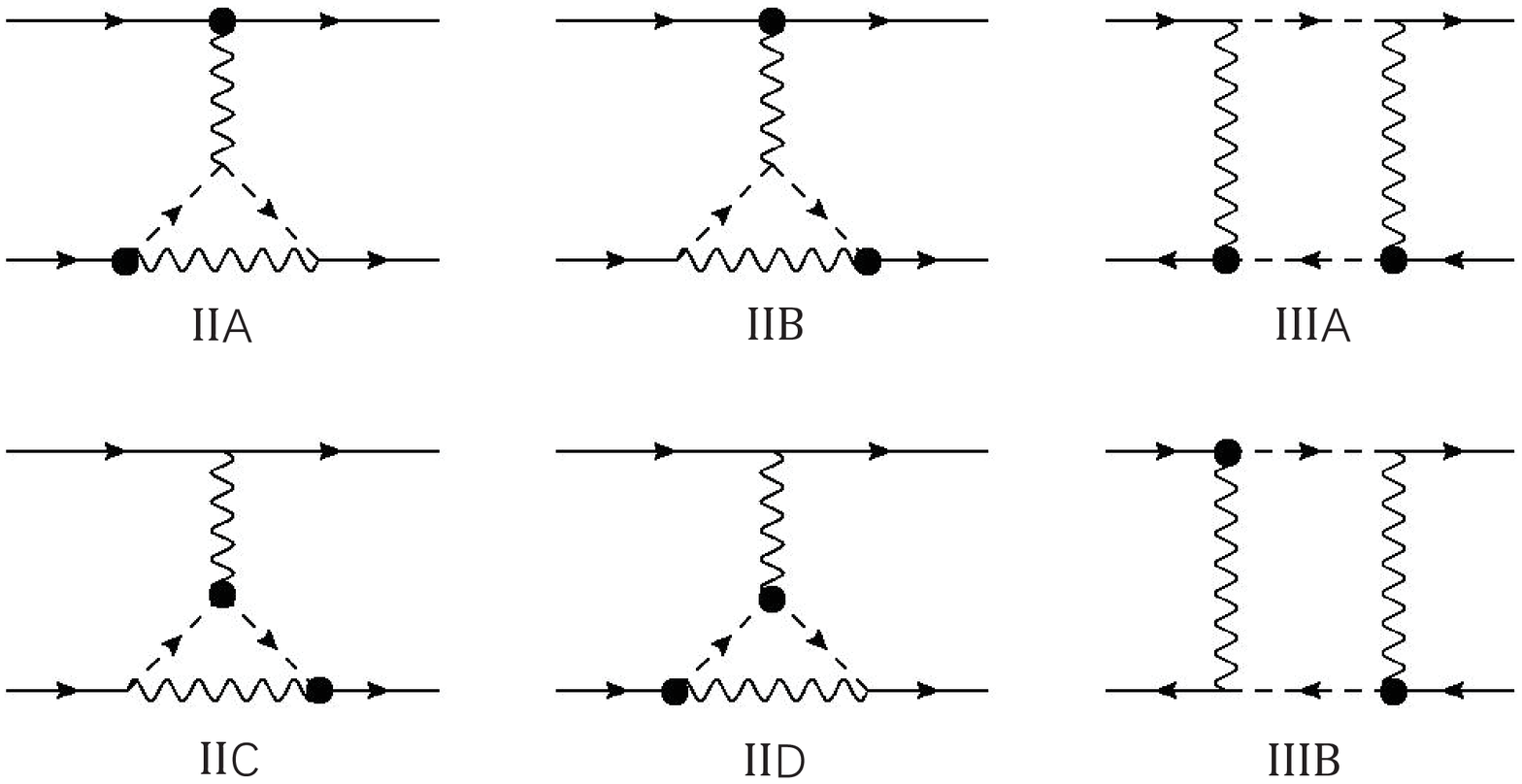}
\caption{(color online). The rest of the particle-particle diagrams that we group as class-II and -III. The dashed lines denote the internal propagators whose momentum lie in the fast mode regime. }
\label{figs3}
\end{figure}

There are four topologically distinct diagrams in Class-II as shown by the IIA, IIB, IIC, and IID diagrams in Fig.\ref{figs3}. After derivation, all of the four diagrams encounter similar integrals of frequency as those in class I, with vanishing integrals. Therefore, up to the leading second order perturbation, we are only left with two particle-hole diagrams, IIIA and IIIB,  as shown by Fig.\ref{figs3}.
Through a direct calculation of the sum of the two diagrams, the most relevant interaction generated in the low-energy is reduced to
\begin{equation}\label{eqn30}
V^{(2)}_{eff}=\frac{v^2_2}{2\Lambda v_F}\sum_{\mathbf{k}_1,\mathbf{k}_2,\mathbf{q}}f^{\dagger}_{a,\alpha,\mathbf{k}_1}f_{a,\alpha,\mathbf{k}_1+\mathbf{q}}f^{\dagger}_{b,\beta,\mathbf{k}_2}f_{b,\beta,\mathbf{k}_2-\mathbf{q}}.
\end{equation}
Therefore, the bare interaction $V^{(2)}_{int}$ generates a leading contribution, $V^{(2)}_{eff}$ to second order calculation, which should dominant in low-energy and determine the possible instabilities, if any.

\section{mean-field study of the s-wave and p-wave excitonic insulator}
Let us focus on the emergent model $H=\sum_{a=\pm}H_{0,a}+V^{(2)}_{eff}$, and temporarily neglect $V^{(1)}_{int}$, where
\begin{equation}\label{eqn31}
  H_{0,a}=a v_F\sum_{\mathbf{k}}f^{\dagger}_{a,\mathbf{k}} \mathbf{k}\cdot\boldsymbol{\sigma}^{(T)}f_{a,\mathbf{k}},
\end{equation}
and
\begin{equation}\label{eqn32}
  V^{(2)}_{eff}=v_{eff}\sum_{\mathbf{k}_1,\mathbf{k}_2,\mathbf{Q}}f^{(\rho)\dagger}_{\mathbf{k}_1,\alpha}
  f^{(\rho)}_{\mathbf{k}_2,\alpha}f^{(\lambda)\dagger}_{\mathbf{k}_2+\mathbf{Q},\beta}f^{(\lambda)}_{\mathbf{k}_1+\mathbf{Q},\beta},
\end{equation}
where we have replaced $\mathbf{q}$ by $\mathbf{Q}$ defined by $\mathbf{Q}=\mathbf{k}_2-\mathbf{k}_1-\mathbf{q}$. The interaction can be further divided into inter- and intra-valley interaction and reduced in the channel with $\mathbf{Q}=0$, as also suggested by the above RG analysis. This generates,
\begin{equation}\label{eqn33}
  V^{intra}_{eff}=v_{eff}\sum_{\mathbf{k},\mathbf{k}^{\prime},\rho}f^{(\rho)\dagger}_{\mathbf{k},\alpha}
  f^{(\rho)}_{\mathbf{k}^{\prime},\alpha}f^{(\rho)\dagger}_{\mathbf{k}^{\prime},\beta}f^{(\rho)}_{\mathbf{k},\beta},
\end{equation}
and
\begin{equation}\label{eqn34}
  V^{inter}_{eff}=v_{eff}\sum_{\mathbf{k},\mathbf{k}^{\prime},\rho}f^{(\rho)\dagger}_{\mathbf{k},\alpha}
  f^{(\rho)}_{\mathbf{k}^{\prime},\alpha}f^{(\overline{\rho})\dagger}_{\mathbf{k}^{\prime},\beta}f^{(\overline{\rho})}_{\mathbf{k},\beta}.
\end{equation}
From the RG analysis, all diagrams except for the particle-hole diagrams IIIA and IIIB vanishes, suggesting a particle-hole instability for large $V^{(2)}_{int}$. We thereby diagonalize $H_0$ to the band basis, whose quasi-particles are represented by the c-fermions. The interaction is transformed into combination of many scattering terms. Keeping the particle-hole scattering with energy conservation, we obtain from $V^{intra}_{eff}$ the following scattering terms:
\begin{equation}\label{eqn35}
  V^{intra}_{s_1}=\frac{v_{eff}}{2}\sum_{\mathbf{k},\mathbf{k}^{\prime},a,n}(c^{\dagger}_{a,n,\mathbf{k}}c_{a,n,\mathbf{k}^{\prime}}c^{\dagger}_{a,\overline{n},\mathbf{k}^{\prime}}c_{a,\overline{n},\mathbf{k}}+c^{\dagger}_{a,n,\mathbf{k}}c_{a,\overline{n},\mathbf{k}^{\prime}}c^{\dagger}_{a,\overline{n},\mathbf{k}^{\prime}}c_{a,n,\mathbf{k}}),
\end{equation}
\begin{equation}\label{eqn36}
  V^{intra}_{s_2}=\frac{v_{eff}}{2}\sum_{\mathbf{k},\mathbf{k}^{\prime},\rho}\cos(\theta-\theta^{\prime})(c^{\dagger}_{a,n,\mathbf{k}}c_{a,n,\mathbf{k}^{\prime}}c^{\dagger}_{a,\overline{n},\mathbf{k}^{\prime}}c_{a,\overline{n},\mathbf{k}}-c^{\dagger}_{a,n,\mathbf{k}}c_{a,\overline{n},\mathbf{k}^{\prime}}c^{\dagger}_{a,\overline{n},\mathbf{k}^{\prime}}c_{a,n,\mathbf{k}}).
\end{equation}
We now perform Hubbard-Stratonovich decomposition by introducing bosonic fields. In mean-field level, the bosonic fields reads as
\begin{eqnarray}
  \chi_{s} &=& v_{eff}\sum_{\mathbf{k}}\langle c^{\dagger}_{a,n,\mathbf{k}}c_{a,\overline{n},\mathbf{k}}\rangle, \\
  \boldsymbol{\chi}_{p} &=&  v_{eff}\sum_{\mathbf{k}}\langle \hat{\mathbf{k}} c^{\dagger}_{a,n,\mathbf{k}}c_{a,\overline{n},\mathbf{k}}\rangle,
\end{eqnarray}
where $\chi_{s,p}$ are the $s$, $p$-wave EI order. The vectorial p-wave order parameter has stable solution with $p+ip$-wave configuration \cite{ruib}, i.e., $\boldsymbol{\chi}_p= \frac{1}{\sqrt{2}}\chi_p(\hat{\mathbf{e}}_x+i\hat{\mathbf{e}}_y)$, where $\chi_p$ is treated as the mean-field order parameter. Inserting the above mean-field orders into the Hamiltonian, a mean-field theory can be obtained. By minimizing the ground state energy, one then derives the following self-consistent equations as,
\begin{eqnarray}
  \overline{\chi}_p &=& \frac{\overline{v}_{eff}}{4\pi}\int^1_0dk^{\prime}k^{\prime}\int^{2\pi}_0d\theta\frac{\overline{\chi}_p+\sqrt{2}\overline{\chi}_s\cos\theta}{\sqrt{\overline{\chi}^2_p/2+\sqrt{2}\overline{\chi}_p\overline{\chi}_s\cos\theta+\overline{\chi}^2_s+(\overline{\lambda}_2+k^{\prime})^2}} , \\
  \overline{\chi}_s &=& \frac{\overline{v}_{eff}}{4\pi}\int^1_0dk^{\prime}k^{\prime}\int^{2\pi}_0d\theta\frac{2\overline{\chi}_s+\sqrt{2}\overline{\chi}_p\cos\theta}{\sqrt{\overline{\chi}^2_p/2+\sqrt{2}\overline{\chi}_p\overline{\chi}_s\cos\theta+\overline{\chi}^2_s+(\overline{\lambda}_2+k^{\prime})^2}},
\end{eqnarray}
where $\theta$ is the angle of $\mathbf{k}$. We have normalized the momentum by the cutoff $\Lambda$, resulting in the integral of a dimensionless limit ``wave vector" $k^{\prime}$ from 0 to 1. The order parameters $\chi_{s,p}$ are also normalized by $v_F\Lambda$ to their dimensionless version, $\overline{\chi}_{s,p}$. The inter-valley mean-field theory describing the CDW instability can be similarly obtained.

\section{global mean-field theory with respect to the CS superconductor and exciton insulator order}
We now consider $V^{(1)}_{int}$ and $V^{(2)}_{eff}$ simultaneously. We introduce bosonic mean-fields to decouple $V^{(1)}_{int}$, which in mean-field level, reads as, $\hat{\Delta}_{\mathbf{k}}\equiv\Delta^{\rho\lambda}_{\mathbf{k}}=\sum_{\mathbf{k}^{\prime}}v^{\alpha\beta\rho\lambda}_{\mathbf{k}-\mathbf{k}^{\prime}}\langle f_{a,\alpha,\mathbf{k}^{\prime}}f_{\overline{a},\beta,-\mathbf{k}^{\prime}}\rangle$, which further generates two mean-field order parameters $\Delta_{3k}$ and $\Delta_{0k}$ \cite{Ruia}.
Then, taking into account the above mean-field theory with respect to $V^{(2)}_{eff}$ in the $s$-wave CS EI channel, we obtain the following two BdG Hamiltonian in the basis $\Psi_{\mathbf{k}}=[f_{a,A,\mathbf{k}},f_{a,B,\mathbf{k}},f^{\dagger}_{\overline{a},A,-\mathbf{k}},f^{\dagger}_{\overline{a},B,-\mathbf{k}}]^\mathrm{T}$,
\begin{equation}\label{eqn37}
  \mathcal{H}_{BdG}=\left(
                        \begin{array}{cccc}
                          \chi_{+,s} & v_F(k_x-ik_y) & \Delta_{3k} & \Delta_{0k}e^{-i\theta} \\
                          v_F(k_x+ik_y) & -\chi_{+,s} & -\Delta_{0k}e^{i\theta} & \Delta_{3k} \\
                          \Delta_{3k} & -\Delta_{0k}e^{-i\theta} & \chi_{-,s} & -v_F(k_x-ik_y) \\
                          \Delta_{0k}e^{i\theta} & \Delta_{3k} & v_F(k_x+ik_y) & -\chi_{-,s} \\
                        \end{array},
                      \right)
\end{equation}
where, to be more general, we introduce $\chi_{\pm,s}$ to denote the EI order for $a=\pm$ valley. Due to the valley symmetry of the mean-field Hamiltonian in the last section, $\chi_{+,s}$ and $\chi_{-,s}$ must have same magnitude but there can allow a difference of minus sign between them, i.e., $\chi_{+,s}=m\chi_{-,s}$ with $m=\pm$.
$\Delta_{3k}$, $\Delta_{0k}$, together with the CS excitonic order $\chi_{a,\pm}$, are the mean-field parameters that need to be self-consistently solved. From the above BdG Hamiltonian, we can obtain the energy dispersion of the lower two bands for both $m=\pm1$, i.e.,
\begin{equation}\label{eqn37}
  E_{m=+}=\sum_{\mathbf{k},a=\pm}-\sqrt{(\Delta_{0k}+akv_F)^2+(\Delta_{3k}-a\chi_s)^2},
\end{equation}
and
\begin{equation}\label{eqn38}
  E_{m=-}=\sum_{\mathbf{k},a=\pm}-\sqrt{(\Delta_{0k}+a\sqrt{\chi^2_s+k^2v^2_F})^2+\Delta^2_{3k}}.
\end{equation}
It is obvious that, for $\Delta_{0k}=\Delta_{3k}=0$, i.e., one turns off the effect of $V^{(1)}_{int}$, the energy of the two configuration $m=+1$ and $m=-1$ are degenerate, as illustrated by the main text. However, the dispersion of the two cases become quite different, with taking into account $V^{(1)}_{int}$ and $\Delta_{0k},\Delta_{3k}\neq0$. A direct comparison of the energies then illustrate the time-reversal symmetry breaking configuration $m=-1$ is favored spontaneously. Therefore, from the above analysis, one can readily observe the TRS breaking, as a result of competitions between $J_1$ and $J_2$.

The self-consistent equation for $m=-1$ is then obtained as,
\begin{eqnarray}
  \Delta_{3k} &=& \frac{v_F}{2}\sum_{a=\pm}\int^{\infty}_kdk^{\prime}\frac{\Delta_{0k^{\prime}}+a\sqrt{\chi^2_s+k^{\prime2}v^2_F}}{E_{a,\mathbf{k}^{\prime}}} \\
  \Delta_{0k} &=& \frac{v_F}{2}\sum_{a=\pm}\int^{k}_0dk^{\prime}\frac{k^{\prime}\Delta_{3k^{\prime}}}{kE_{a,\mathbf{k}^{\prime}}}, \\
  \chi_s &=& \frac{v_{eff}}{\Lambda^2}\sum_{a=\pm}\int^{\Lambda}_0dkk\frac{a\Delta_{0k}+\sqrt{\chi^2_s+k^2v^2_F}}{E_{a,\mathbf{k}}\sqrt{\chi^2_s+k^2v^2_F}}\chi_s,
\end{eqnarray}
where $E_{a,\mathbf{k}}=-\sqrt{(\Delta_{0k}+a\sqrt{\chi^2_s+k^2v^2_F})^2+\Delta^2_{3k}}$. In the long-wave regime $k\rightarrow0$, we obtain $\Delta_{0k}=v_Fk\gamma/2$, and $\Delta_{3k}=\Delta_3$ being independent of momentum, where $\gamma=\Delta_3/\sqrt{\chi^2_s+\Delta^2_3}$. Thus, in the long-wave length regime, we only need to self-consistently solve $\Delta_3$, $\chi_s$.
After introducing dimensionless quantities as $\overline{k}=k/\Lambda$, $\overline{\Delta}_3=\Delta_3/\Lambda v_F$, $\overline{\chi}_s=\chi_s/\Lambda v_F$, and $\overline{v}_{eff}=v_{eff} /ev_F\Lambda$, the self-consistent equations with respect to $\Delta_3$, $\chi_s$ are reduced into:
\begin{eqnarray}
  \overline{\Delta}_3 &=& \frac{e}{2}\int^{1}_0 d\overline{k}\frac{e\overline{k}\gamma/2+\sqrt{\overline{\chi}^2_s+\overline{k}^{2}}}{\sqrt{(e\overline{k}\gamma/2+\sqrt{\overline{\chi}^2_s+\overline{k}^{2}})^2+\overline{\Delta}^2_3}}+
  \frac{e\overline{k}\gamma/2-\sqrt{\overline{\chi}^2_s+\overline{k}^{2}}}{\sqrt{(e\overline{k}\gamma/2-\sqrt{\overline{\chi}^2_s+\overline{k}^{2}})^2+\overline{\Delta}^2_3}}, \\
  \overline{\chi}_s &=& e\overline{v}_{eff}\int^{1}_0d\overline{k}\overline{k}
  \{\frac{\overline{k}\gamma/2+\sqrt{\overline{\chi}^2_s+\overline{k}^{2}}}{\sqrt{[(\overline{k}\gamma/2+\sqrt{\overline{\chi}^2_s+\overline{k}^{2}})^2+\overline{\Delta}^2_3][\overline{\chi}^2_s+\overline{k}^{2}]}} +\frac{-\overline{k}\gamma/2+\sqrt{\overline{\chi}^2_s+\overline{k}^{2}}}{\sqrt{[(\overline{k}\gamma/2-\sqrt{\overline{\chi}^2_s+\overline{k}^{2}})^2+\overline{\Delta}^2_3][\overline{\chi}^2_s+\overline{k}^{2}]}}\}\overline{\chi}_s.
\end{eqnarray}
As shown by the main text, the analytic derivation above establishes a parton  mean-field theory based on CS fermions, which can describe the UPT from a N\'{e}el AFM order to a non-uniform CSL in mean-field level.




\date{\today }
\maketitle

\end{document}